\documentclass[twocolumn,aps,prl,showpacs]{revtex4}
\usepackage{ae} 
\usepackage[T1]{fontenc}
\usepackage[ansinew]{inputenc}
\usepackage{amsmath}
\usepackage{graphicx}
\usepackage{color}
\usepackage[colorlinks]{hyperref}

\begin{document}

\title{Magnetic breakdown induced Peierls transition}
\author{A. M. Kadigrobov $^{1,2}$, A. Bjeli\v{s}$^{2}$, and D. Radi\'{c}$^{2}$ }
\affiliation{$^{1}$Theoretische Physik III, Ruhr-Universit\"{a}t Bochum, D-44801 Bochum,
Germany\\
$^{2}$ Department of Physics, Faculty of Science, University of Zagreb, POB 162, 10001 Zagreb, Croatia}
\date{\today}

\begin{abstract}
We predict  the new type of phase transition in quasi one-dimensional system of interacting electrons at high magnetic fields, the stabilization
of a density wave which transforms a two dimensional open Fermi surface into a periodic chain of large pockets with small distances between
them. We show that quantum tunneling of electrons between the neighboring closed orbits enveloping these pockets transforms the electron
spectrum into a set of extremely narrow energy bands and gaps that decreases the total electron energy, thus leading to a \emph{magnetic
breakdown induced density wave} ground state analogous to the well-known instability of Peierls type.


\end{abstract}
\maketitle

\begin{figure}
\centerline{\includegraphics[width=6.0cm]{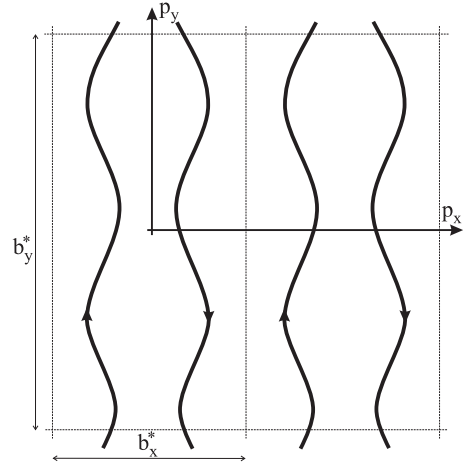}} \caption{ Quasi one-dimensional Fermi surface, with arrows denoting directions of
semiclassical electron motion under a magnetic field perpendicular to the plane $(p_x,p_y)$. $b_{x}^{*}$ and $b_{y}^{*}$ are reciprocal lattice
constants.} \label{1}
\end{figure}

Interacting electrons with quasi one-dimensional (Q1D), i. e. open and corrugated, Fermi surfaces show in many cases instabilities leading to a
density wave (DW) ground state with a broken translational symmetry \cite{reviewDW}. This type of phase transition, predicted by Peierls in 1955
\cite{peierls}, occurs due to divergent charge fluctuations which open the gap at the Fermi surface and so lower the energy of the new ground
state. While the dominance of a repulsive Coulomb or an attractive phonon mediated electron-electron interaction are as a rule responsible for
respective spin or charge modulation of DW, its stabilization depends on the degree of nesting between the left and right two dimensional Fermi
surfaces shown in Fig. \ref{1}. Nesting in real materials is always imperfect, leading to two-dimensional \cite{comment1} pockets of finite size
after determining the optimal wave vector of DW ordering, ${\bf Q} = (2k_F + \delta k_\parallel, b^{*}/2 + \delta k_\perp)$, where $k_F$ and
$b^{*}$ are Fermi wave number and transverse reciprocal lattice vector respectively, while $\delta k_{\parallel,\perp}$ are small corrections.
Unlike the nested parts of the Fermi surface, these pockets develop a weaker gap in the electron spectrum, or even remain gapless. The pockets
thus act against the DW, and, if large enough, may completely eliminate it.

 A qualitative change however takes place after applying strong
enough magnetic field $\textbf{H}$ perpendicular to the plane ($p_{x}, p_{y}$) in Fig. \ref{1}. Landau quantization of band states then reduces
the transverse delocalization of electrons, making the system "more one-dimensional" \cite{GorLeb} and therefore more susceptible to the DW
ordering. DWs are then strengthened, or even newly established in systems which otherwise remain metallic down to $T=0$. The latter case of
field-induced density wave (FIDW) is particularly relevant for our further considerations. The best known examples are those realized in a
series of Bechgaard salts with spin modulation \cite{reviewFISDW}, in which the FIDW order takes place in available magnetic fields up to $\sim
50$T, provided the deviation form perfect nesting is not large. The pockets with discrete Landau orbits are then small and, after establishing
DW gap in dominant "nesting" parts of Fermi surface, quite distant one from another.

The Landau quantization leads to the localization of electron motion, therefore increasing the band energy in general, or leaving it unchanged
for the discrete set of values of $\textbf{H}$ for which the Fermi energy lays exactly in the middle between two neighboring Landau levels
\cite{dHvA}. Still, it contributes to the FIDW stabilization through an additional gain in the correlation energy part due to the improved
longitudinal localization of electron states. On the other hand, the magnetic breakdown effect \cite{Cohen,Slutskin}, i. e. the magnetic field
assisted electron tunneling through barriers between neighboring pockets which delocalizes the electron motion and so decreases the band energy,
is far too weak to be relevant.

At this point we come to our central question, namely, is it possible to stabilize a density wave order with the help of orbital quantization in
the quasi one-dimensional band, but through a predominant decrease of the band energy due to the magnetic breakdown? This \emph{magnetic
breakdown induced density wave} (MBIDW) is possible only by strengthening the tunneling between neighboring pockets, i. e. by choosing the DW
wave vector which, oppositely to FIDW in Bechgaard salts, favors short and weak tunnel barriers together with large pockets between them. In the
present work we examine this new possibility by using the standard mean field approach for DW instability \cite{Gruner} and the semiclasical
treatment of MB tunneling \cite{Slutskin}, and establish the conditions under which this gain overwhelms the elastic energy loss due to the
lattice deformation \cite{comment2}, leading so to the MBIDW stabilization.

\begin{figure}
\centerline{\includegraphics[width=8.6cm]{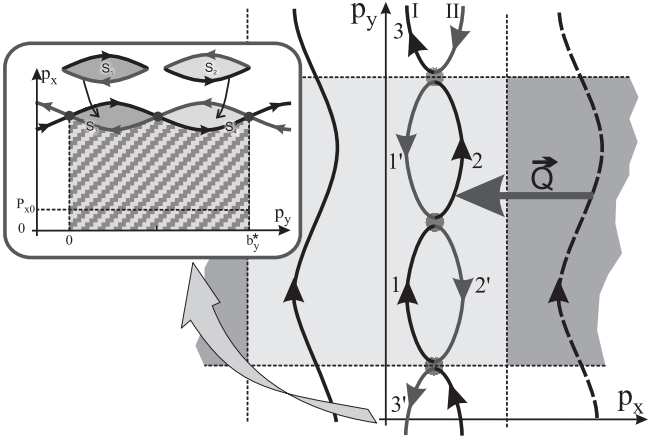}} \caption{(a) Periodic net of large closed orbits separated by small classically
inaccessible MB arrears (thick dots). $\vec{Q} =(Q \approx 2k_F,0)$ is lattice deformation wave vector, and the distance between closed orbits
is proportional to the product of electron mass and deformation potential $V(x)$. (b) Areas of closed orbits, $S_{1,2}$, and areas under open
orbits, $S_{3,4}$ for the respective left and right motion of electrons, determine the phases of semi-classical electron wave functions defined
in the text.} \label{2}
\end{figure}


{\it Qualitative considerations.} Let us choose the deformation wave vector $\textbf{Q}=(Q\approx 2k_F, 0)$ creating a static periodic lattice
distortion and a charge modulation $V(x)$ with the period $2\pi \hbar/Q$ that combines open trajectories with opposite directions of the
electron motion, shown by arrows in Fig. \ref{2}. We encounter a chain of closed orbits encircling large pockets and small areas between them
(shown with thick dots) at which MB takes place.  Here we treat MB electron dynamics within the semi-classical approximation: an electron moving
along a semi-classical section of the chain, say 1, is scattered at the MB area to section 2 and section 2' with the amplitude probabilities
$\rho$ and $\tau$ respectively, fulfilling the normalization condition $|\rho|^2+|\tau|^2=1$ with the MB probability given by
\begin{equation}
|\rho|^2 =\exp(-\gamma).
 \label{r}
\end{equation}
where $\gamma\equiv \pi \frac{|V_0|^2}{ \sigma |v_x v_y|}$. Here
$v_x$ and $v_y$ are longitudinal and perpendicular projections of
the electron velocity $\vec{v} =\partial \varepsilon
(\vec{p})/\partial \vec{p}$ at the point of the magnetic
breakdown, $\sigma =\hbar e H/c$, $e$ and $c$ are the electron
charge and the light velocity, respectively, while $V_0$ is the
non-diagonal matrix element of $V(x)$. For the electron-phonon
system within the mean field approximation, $|V_0|=2g b$ where $g$
is the coupling constant and $b =<b_{Q}>$ is the mean value of the
phonon annihilation operator at the wave number $Q$ (see e.g.
\cite{Gruner}), here directly proportional to the amplitude of the
lattice displacement. Thus, due to the finite lattice distortion
$V(x)$ electron scatters in a periodic set of regions of MB,
latter having a role analogous to that of atomic lattice in an
one-dimensional metal.

\begin{figure}
\centerline{\includegraphics[width=8.0cm]{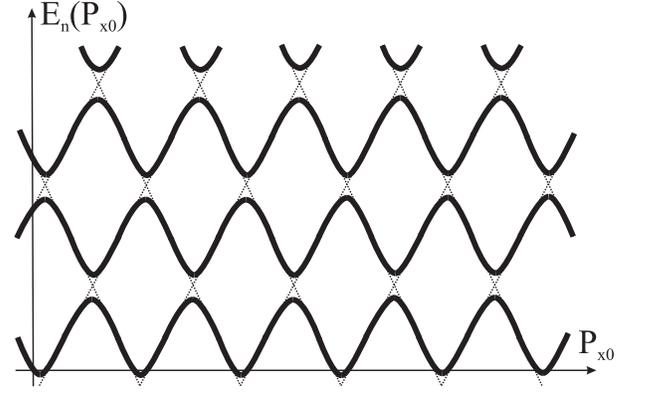}} \caption{Energy spectrum for electrons shown in Fig. \ref{2} in the regime of strong MB
($|\tau|^2\ll 1$). $n$ is the MB band number and $P_{x0}$ is the conserved generalized electron momentum. The period of $E_n(P_{x0})$ is $2\pi
\sigma /B_yv$, while the widths of the energy bands and gaps are $\Delta E_{band} \sim |\rho|^2 \hbar \omega_c$ and $\Delta E_{gap} \sim
|\tau|^2 \hbar \omega_c$ respectively ($\omega_c$ is the cyclotron frequency). Dotted straight lines denote the electron spectrum in the absence
of MB.}

\label{3}
\end{figure}

In order to understand qualitatively the structure of the energy
spectrum of an electron on the orbit chain let us assume at first
that $|V(x)|$ is negligibly small while the strength of the
magnetic field is finite. In this case the probability of MB is
$|\rho|^2=1$ and the electron moves along an open trajectory 1-2-3
... or 1'-2'-3'... (trajectories I and II, respectively). It is
known \cite{1dspectrum} that the electron spectrum
$E^{I,II}_n(P_{x0})$ of such motion is continuous and
characterized by a discrete quantum number $n$ and a continuous
momentum $P_{x0}$. Here superscripts refer to trajectories I and
II; $P_{x0}$ is the generalized momentum projection conserved in
the  gauge with the vector potential $\vec{A}=(-Hy,0,0$). The
average quantum-mechanical electron velocities are $v^{I,II}=d
E^{I,II}_n(P_{x0})/d P_{x0}$ ($v^{I}>0$ and $v^{II}< 0$). This electron spectrum is represented by the dotted straight lines in Fig. \ref{3}.

For a finite charge modulation $V(x)$, the MB is not total ($|\rho|^2 <1$) and the trajectories I and II are coupled  by the finite tunneling
probability. The degeneracy at the crossing points in the electron spectrum is then lifted by $|\tau|^{2} $, producing there energy gaps
analogous to those in the spectrum of a real one-dimensional electron gas created by the charge modulation $V(x)$ itself \cite{mingap}.

If the Fermi energy occurs in the middle of one of these gaps, that is at one of the crossing points in the unperturbed spectrum, the electron
band energy decreases. This energy gain grows by increasing the gap, which in turn increases with $|\tau|$. As $|\tau|$ increases by increasing
$|V_0|$ as is seen from Eq. \ref{r}, the energy band gain is favored by the increase of the charge modulation $V(x)$. However, by increasing
$V(x)$ one increases the lattice energy and, as a result, competition of these two tendencies, a decrease of the band energy and an increase of
the elastic one, gives the optimal value of the charge modulation $V(x)$.

{\it Analytical calculations.}   In the $\vec{p}$-space the
electron moves along  classical trajectories $p^{(l)}_x(p_y )$
(marked by $l= 1,1^{'},...$ in Fig. \ref{2}) between the points of
MB. Corresponding semi-classical wave functions are
\begin{equation}
a^{(l)}\exp\{-\frac{i}{\sigma}\int_{}^{p_y} [p_{x}^{(l)}(\bar{p}_y)-P_{x0}]d\bar{p}_y\},
\label{WF}
\end{equation}
where $p^{(l)}_x(p_y )$ are solutions of the equation $\varepsilon_s(p_x,p_y)=E$ while $\varepsilon_s(\vec{p})$ is the electron dispersion law
for a finite $V(x)$ and at $H=0$, $s =1,2$ indexing the new electron bands. At the points of magnetic breakdown the in-coming and out-coming
functions ({\ref{WF}) are coupled by the magnetic breakdown $2\times2$ unitary matrix \cite{SK,Slutskin,S}
\begin{equation}
\hat{\tau} =e^{i\Psi}\left(\begin{matrix} |\tau|e^{i \eta}& |\rho|e^{i \nu}\\
-|\rho|e^{-i \nu } & |\tau|e^{-i\eta}
\end{matrix} \right),
\label{taumatrix}
\end{equation}
e. g. $a_2= \exp(i\Psi)(a_1 |\rho|\exp(i\nu)+a_{1^{'}}|\tau|\exp(i
\eta))$. The  phases depend on the MB parameter $\gamma$ from Eq.
\ref{r}. In the limit $\gamma\ll 1$ (that is $|\tau|^2\ll 1$) to
which we limit our calculations of MBIDW, $\eta \approx -\pi/4
-\gamma \ln \gamma $, $\nu \approx 0, \Psi \approx 0$.

Matching  wave functions (\ref{WF}) at the MB points with matrix
(\ref{taumatrix}) and taking into account the electron phase gains
between them one finds the dispersion function
\begin{equation}
D(E,P_{x0}) = \cos \bar{\Phi}_{-} -|\tau|^2\cos\Phi_{+}- |\rho|^2\cos(\Phi_0  -\frac{P_{x0}b^{*}_y}{\sigma}).\label{D}
\end{equation}
Here $\bar{\Phi}_{-}=\Phi_{-}+2\Psi$, $\Phi_{\pm} =(1/2\sigma)(S_1(E)\pm S_2(E))$ and $\Phi_0 =(1/2\sigma) (S_3(E)-S_4(E))+2\nu$, while
$S_{1,2,3,4}(E,Q)$ are the areas enclosed by electron trajectories which depend on the position of the crossing points of trajectories I and II
determined by $\vec{Q}$ (see Fig. \ref{2}). The electron spectrum is determined by dispersion equation $D(E,P_{x0})=0$.

If $|\rho|^2=0$,  this equation gives a set of discrete Landau levels, while in the case $|\rho|^2  \neq 0$ the degeneracy with respect to
$P_{x0}$ is lifted and the electron spectrum depends on a continuous quantum number, $E=E_n(P_{x0})$, where the band number $n$ enumerates
solutions of the dispersion equation. Solving this equation in the limit $|\tau|^2\ll 1$ one finds the electron spectrum presented in Fig.
\ref{3}.

In order to find an explicit formula  for the  density of states (DOS) for electrons moving along the MB chain, we express it in terms of the
spectral function (\ref{D}) as follows:
\begin{eqnarray}
\nu_{MB} (E)=\int \sum_n \delta (E-E_n (P_{x0}))
\frac{dP_{x0}}{2\pi \hbar}\nonumber \\
=\int |D^{'}(E,P_{x0})|\delta(D(E,P_{x0}))\frac{dP_{x0}}{2\pi \hbar},
 \label{densityintermidient}
\end{eqnarray}
where here and below $f^{'} \equiv \partial f/\partial E$.  Integrating with respect to $P_{x0}$ we find
\begin{eqnarray} \label{density}
\nu_{MB} (E) & = & \frac{\sigma}{(2\pi)^2}\Theta \left(|\rho|^2-\left|
 \cos\Phi_{-} - |\tau|^2 \cos\Phi_{+}\right|
\right) \nonumber \\
& & \times \frac{|\phi^{'}_{-}\sin\Phi_{-}-|\tau|^2 \phi^{'}_{+}\sin\Phi_{+}|}{\sqrt{|\rho|^4-\left( \cos\Phi_{-} - |\tau|^2
\cos\Phi_{+}\right)^2}},
\end{eqnarray}
where $\Theta (x)$  equals unity for $x \geq 0$ and zero
otherwise. From here one easily sees that at $|\tau|^2=0$ (that is
at $V_0=0$) DOS is equal to the one in the absence of MBIDW .

Eq. \ref{density} permits to find the number of electrons $N=\int [\exp(E-\mu)/T +1 ]^{-1}\nu(E)dE$ and the electron free energy $F= N \mu -T
\int \ln [1+ \exp(\mu -E)/T ]  \nu(E) dE$, $\mu$ being the chemical potential. We calculate them expanding the right-hand side of Eq.
\ref{density} in a double Fourier series in $\Phi_{\pm}$ and in a power series in $|\tau|^2$. Below we solve the problem in the vicinity of the
critical temperature $T_c$ in which the order parameter and the potential $V_0$ are small, so that $|\tau|^2 \ll 1$ and we can keep only terms
of the lowest order ($\sim|\tau|^2$) in the above-mentioned power series.

After finding $N$ and $F$ we take into account the conservation of electron number and determine the change of the chemical potential $\delta
\mu =\mu_{MB} - \mu_0$, where $\mu_{MB}$ and $\mu_0$ are chemical potentials at $V(x)\neq 0$ and $V(x)=0$ respectively. Inserting it in the free
energy and neglecting terms of the order of $|\tau|^4\hbar \omega_c/\varepsilon_F $ and $\tau^2(\hbar \omega_c/\varepsilon_F )^2$, we find the
difference of the total free energies in the presence and the absence of MBIDW at $T>\hbar\omega_c^{(+)}/\pi$, where $\omega_c^{(+)}=eH/m^*_+ c$
with the effective mass given by $m^*_+= (m^*_1+m^*_2)/2$ and $m^*_{1,2}\equiv|\partial (S_{1,2})/\partial \varepsilon |$:
\begin{eqnarray} \label{totalFE}
\Delta F & = & \Delta F_0 \left[1-\exp\left(-\frac{\pi g^2 b^2}{
\sigma|v_x v_y|}\right)\right]
\cos\frac{S_+}{2\sigma}\sin\frac{S_-}{2\sigma}\nonumber \\
 & & + b^2 \hbar \omega_{Q};\hspace{0.4cm}
 \Delta F_0 \equiv kT\exp\{-\frac{\pi
 T}{\hbar\omega_c^{(+)}}\}.
\end{eqnarray}
Here   $S_{\pm}= S_1(\mu_0,Q)\pm S_2(\mu_0,Q)$ is the effective electron loop area, and the last term is the lattice elastic energy given by
phonon frequency $\omega_{Q}$ at momentum $Q$. Eq. \ref{totalFE} is valid in the regime  $\sigma \lesssim |S_-|\ll S_+$.

From Eq. \ref{totalFE} it follows that $\Delta F$ has a series of minima at the discrete set of lattice deformation wave numbers $Q_n$ for which
$S_-(\mu_0,Q)/2\sigma= (\pi/2)(2n+1)$ with integer values of $n$ (for $\partial S_1/\partial Q \approx -\partial S_2/\partial Q$)
\cite{comment3}. On the other hand, the minimization of Eq. \ref{totalFE} with respect to $b^2$ shows that $\Delta F $ is negative and has an
absolute minimum at
\begin{eqnarray}
b^2=\frac{\sigma|v_x v_y|}{\pi g^2 }
\ln\left[\lambda \frac{\mu_0 \left|\cos(S_{+}/2\sigma)\right| \pi T}{m^{*}_{+}|v_x v_y| \hbar
\omega_c^{+}}\exp\left(-\frac{\pi T}{\hbar\omega_c^{(+)}}\right)\right]
 \label{P}
\end{eqnarray}
as shown in Fig. \ref{4}. Here $\lambda = g^2/(\hbar \omega_{Q_n}\mu_0)$ is the dimensionless electron-phonon coupling constant (see
\cite{Gruner}). The corresponding critical temperature for the transition to the MBIDW state is
\begin{equation}
T_c \approx \frac{\hbar \omega_c^{+}}{\pi }\ln\Bigl({\lambda \frac{\mu_0 \left|\cos(S_{+}/2\sigma)\right|}{m^{*}_{+}|v_x v_y|}}\Bigr)
\label{Tc}
\end{equation}
provided $\lambda \mu_0 \left|\cos(S_{+}/2\sigma)\right| /m^{*}_{+}|v_x v_y|>1 $.

\begin{figure}
\centerline{\includegraphics[width=8.0cm]{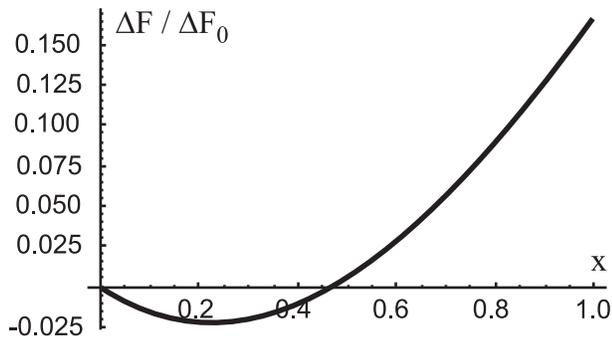}} \caption{The free energy difference $\Delta F/\Delta F_0$ vs. $x \equiv (\pi g^2/ \sigma
|v_x v_y|) b^2$ for $\gamma \equiv (\sigma |v_x v_y|/\pi g^2)(\hbar \omega_{Q}/\Delta F_0)=0.8$ , showing stable minimum of DW ground state at
$x_0 \approx 0.22$.} \label{4}
\end{figure}

In conclusion, we have shown that the MBIDW instability may occur in quasi-one-dimensional conductors in a wide range of parameters and at any
finite magnetic field. The topological prerequisite for such instability is the open Fermi surface of the type shown in Fig. \ref{1}, which
after a periodic breaking of symmetry from Fig. \ref{2}, leads to the lattice of closed orbits connected by barriers at cross points. It was
already pointed out that magnetic field assisted tunneling through these barriers is crucial for the stabilization of MBIDW. MB  introduces a
qualitative change in electron dynamics, transforming the continuous spectrum of quasi one-dimensional interacting electrons under a strong
magnetic field into a set of alternating narrow energy bands and energy gaps, with the widths proportional to cyclotron energy multiplied by
electron MB tunneling and scattering probability respectively. Just a mere Landau quantization of such closed electron orbits does not lead to
the DW stabilization since it only rises the band energy due to the localization of electron motion. Only by delocalization of motion introduced
by MB, whose negligence was probably the reason why this effect was not predicted in preceding literature, the total energy is lowered enough to
grant the stability of DW ground state.

We have also shown that the free energy of MBIDWs has local minima at a set of discrete values of the deformation wave vector $Q$, indicating
the role of resonances between neighboring orbits with generally different size (Fig. \ref{2}). We remind that this series of wave vectors
belong to the "anti-nesting" regime, which again reflects the essential differences between MBIDWs and standard FIDWs, the latter being the
consequence of logarithmic anomalies in the DW correlation function due to the almost perfect nesting between left and right Fermi surfaces in
Fig. \ref{1}. The dependence of $T_c$ on $H$ has a periodic (in $1/H$) set of gaps for a given $Q_n$ (Eq. \ref{Tc}), in contrast to the
corresponding dependence for FIDWs \cite{Lebed}. This, together with a qualitatively different dependence on the coupling constant and band
parameters, could be one guide in the search for MBIDWs. Another one are fast magneto-resistance oscillations due the formation of relatively
large closed orbits (Fig.\ref{2}). The  regime wider than that of Eq. \ref{totalFE} and the peculiarities of MBIDW phase diagram will be
analyzed in a more detailed presentation.

\textbf{Acknowledgements.} A.K. gratefully acknowledges the hospitality of the University of Zagreb. The work is supported by project
119-1191458-1023 of Croatian Ministry of Science, Education and Sports.


\begin{thebibliography}{99}
\bibitem{comment1} The band dispersion in the third spatial dimension is usually
irrelevant and is here neglected for simplicity.
\bibitem{comment2} We adopt here the electron-phonon picture. The analysis can be
straightforwardly extended to the electron-electron one.
\bibitem{mingap}In the case under consideration $|\tau|^2$ plays the role of the
potential, and  the minimal gap is proportional to $|\tau|^2$ that is the minimal gap is $\sim |\tau|^2\hbar \omega_c$.
\bibitem{comment3} Calculations  show that at $T \to 0$ one has $S_{-}/2 \sigma \to 2 \pi n$ that corresponds to the position
of $\mu_0$  inside one of the energy gaps of Fig. \ref{3} (see Eq. \ref{density}).
\end{thebibliography}
\end{document}